# Formation of Epitaxial Graphene on SiC(0001) using Vacuum or Argon Environments

Luxmi, N. Srivastava, and R. M. Feenstra
Department of Physics, Carnegie Mellon University, Pittsburgh, PA 15213
P. J. Fisher
IBM T. J. Watson Research Center, Yorktown Heights, NY 10598

**Abstract**
The formation of graphene on the (0001) surface of SiC (the Si-face) is studied by atomic force microscopy, low-energy electron microscopy, and scanning tunneling microscopy/spectroscopy. The graphene forms due to preferential sublimation of Si from the surface at high temperature, and the formation has been studied in both high-vacuum and 1-atm-argon environments. In vacuum, a few monolayers of graphene forms at temperatures around 1400°C, whereas in argon a temperature of about 1600°C is required in order to obtain a single graphene monolayer. In both cases considerable step motion on the surface is observed, with the resulting formation of step bunches separated laterally by ≥10 μm. Between the step bunches, layer-by-layer growth of the graphene is found. The presence of a disordered, secondary graphitic phase on the surface of the graphene is also identified.

## I. INTRODUCTION

Graphene, a single sheet (or a few sheets) of carbon atoms arranged in a hexagonal arrangement, is looked upon as a promising future material for electronic devices due to its unusual properties.[1,2] The origin of these superior properties of graphene lies in its band structure which shows a linear dispersion near Dirac points.[3,4,5] However, the challenge encountered by all researchers is the formation of uniform and homogenous wafer-size graphene films. The method of exfoliating graphene from graphite gives rise to sub-millimeter size graphene flakes leaving this method unsuitable for industrial applications.[1] Thus far, annealing hexagonal polytypes of SiC in vacuum, with the Si atoms preferentially sublimating, seems to be the best available approach to produce graphene films viable for circuit applications.[4,6,7,8] Recent use of alternate environments such as 1 atm of argon or $1\times10^{-8} - 1\times10^{-6}$ Torr of disilane has demonstrated improved quality of the graphene films, at least for the (0001) face of SiC (the so-called Si-face).[9,10,11]

A number of prior works have studied graphene formation on the Si-face of SiC, with scanning tunneling microscopy/spectroscopy (STM/STS), atomic force microscopy (AFM), and low-energy electron microscopy (LEEM) providing detailed real-space views of the process. Pits are formed during the initial stage, associated with the formation of the 6√3×6√3-R30° surface reconstruction.[12,13,14,15] This reconstruction persists as the graphene forms, located at the interface between the graphene and the SiC and thus acting as a template for the graphene formation. LEEM provides a direct measure of the thickness of a graphene film at each point on the surface,[16] revealing oscillations in the reflectivity of electrons which are dependent on the film thickness and beam energy. The number of graphene layers equals the number of minima in the reflectivity curve. Pits are also observed for thicker graphene layers, but in this case they are believed to arise from strain effects due to the difference in thermal expansion coefficients between graphene and SiC.[12]

Graphene films prepared in vacuum show areas with different thicknesses, and hence these films present a problem in circuit applications since the electronic properties are thickness dependent. In order to make wafer scale graphene films with uniform thickness, Virojanadara *et al.* first reported on the formation of homogenous large area graphene films on SiC(0001) surface by annealing at a temperature of 2000°C in a 1-atm argon environment.[9] Emtsev *et al.* also reported on the formation of thin and uniform graphene on the Si-face of SiC by annealing at 1650°C under 900 mbar of argon. At a fixed temperature the Si sublimation rate is significantly decreased in the presence of argon, so to achieve a given thickness of graphene (in a given time) the annealing temperature must be increased. The higher temperature leads to an improved growth of graphene in terms of its homogeneity and thickness. Tromp and Hannon studied the pressure-temperature phase transformations for graphene films prepared on the Si-face of SiC in a background pressure, ranging over $10^{-8} - 10^{-6}$ Torr, of disilane.[11] They also found an increase in the domain size of graphene with uniform thickness.

In this work, we use AFM, LEEM and STM/S to study the morphology of graphene films prepared on the Si-face of SiC. Our results are consistent with the prior works mentioned above, but go beyond those in terms of the combination of *surface* morphology (by AFM) and *graphene* morphology (by LEEM) studied over a range of preparation conditions. For annealing in vacuum, graphene films with thickness < 2 monolayers (ML) reveal many small domains with ≈100 nm lateral extent. With further annealing these domains coarsen to form larger regions of uniform graphene. For films of > 2 ML thickness, steps on the surface (due to unintentional miscut of ≈0.1°) are found to form step bunches, separated laterally by ≳10 μm. On these surfaces the graphene is found to be thicker at the step bunches than between them, as revealed by LEEM. Between the step bunches the *variation* in graphene thickness is limited mainly to 1 ML, indicative of layer-by-layer formation of the graphene, with typical lateral domain sizes of 3 – 4 μm. We also observe what we believe to be a secondary phase on the surface arising from excess carbon in the form of nano-crystalline graphite (NCG). We have furthermore investigated the effect of an argon annealing environment, rather than vacuum, on the morphology and homogeneity of the graphene films. For these samples, the higher temperatures employed for the annealing cause step bunching even for graphene coverages of 1 ML or less. We find no difference in the graphene coverage at the step bunches compared to between bunches, and we achieve domains of ≫10 μm lateral extent with single-ML graphene thickness.

## II. EXPERIMENT

Our experiments are performed on nominally on-axis, *n*-type 6H-SiC or semi-insulating 4H-SiC wafers purchased from Cree Corp., with no apparent differences between results for the two types of wafers. The wafers are 2 or 3 inches in diameter, mechanically polished on both sides and epi-ready on the (0001) surface. The wafers are cut into 1x1 cm$^2$ samples and the samples are chemically cleaned in acetone and methanol before putting them into our custom built preparation chamber which uses a graphite strip heater for heating the samples.[14] Samples are first etched in a 10 lpm flow of pure hydrogen for 3 min at a temperature of 1600°C. This H-etching removes the polishing scratches which arise during the mechanical polishing of the wafers, resulting in an ordered step-terrace arrangement on the surface which is suitable for graphene formation.[17] Before annealing, hydrogen is pumped away from the chamber and we wait until a desired pressure of $10^{-8}$ Torr is reached and then the samples are annealed for 10 to 40 min at temperatures ranging from 1100-1500°C. Temperature is measured using a disappearing filament pyrometer, with calibration done by using a graphite cover over the sample and measuring its temperature.[14]

Following graphitization the samples are transferred to an Elmitec LEEM III system. Samples are outgassed at 700°C prior to study. The sample and the electron gun are kept at a potential of − 20 kV and LEEM images are acquired with electrons having energy, set by varying the bias on the sample, in the range of 0 – 10 eV. The intensity of the reflected electrons from different regions of the sample is plotted as a function of the beam energy. The reflectivity curve shows oscillations, with the number of graphene monolayers (ML = 38.0 carbon atoms/nm$^2$) being given by the number of minima in the curve.[16]

The vacuum system containing the LEEM is also equipped with a 5 kV electron gun and VG Scientific Clam 100 hemispherical analyzer used for Auger electron spectroscopy (AES). For routine determination of graphene thickness by AES we use the ratio of the 272 eV KLL C line to the 1619 eV KLL Si line. This ratio is analyzed with a model involving the escape depths of the electrons,[18] with the overall magnitude of the ratio being calibrated to graphene thicknesses determined by LEEM. For the Si-face, the 6√3×6√3-R30° "buffer" layer between the graphene and the SiC is taken to contain 1 ML worth of carbon,[15] and this amount is subtracted from the carbon coverage determined from AES to yield the graphene coverage values given below. The surface morphology of our graphene films were studied by AFM using a Digital Instruments Nanoscope III in tapping mode.

## III. RESULTS AND DISCUSSION
### A. Graphene Formation in Vacuum

As an introduction to our results for preparation in vacuum, Fig. 1 displays AFM images from two samples that we have studied in detail – one sample prepared at 1320°C for 10 min and with 1.9 ML average graphene coverage, and the other prepared at 1320°C for 40 min and with 3.0 ML graphene coverage. The morphology of these samples is similar to that described in our previous works: After H-etching the SiC surfaces consist of flat terraces separated by single-unit-cell high steps, with the steps arising from unintentional miscut of the wafer. With modest heating in vacuum, pits start to form on the terrace and the steps edges become irregular. Those morphological changes have been explained by Hannon and Tromp as resulting from the formation of the 6√3×6√3-R30° reconstruction on the surface.[15] With more extensive heating the steps rearrange, as in Fig. 1(a), and additional small pits form on the surface.[14,19] Upon further heating, considerable motion of the steps occurs and the surface transforms into one with quite large, ≳10 μm, terraces separated by step bunches; Fig. 1(b) is an image from one such terrace. Images of the terraces sometimes reveal faint networks of "snowflake-like" patterns, as seen in Fig. 1(b). As discussed in detail in the following Section, we believe these patterns to arise from excess carbon on the surface forming disordered, nano-crystalline graphite (NCG). This presence of this NCG is found *not* to significantly interfere with the LEEM imaging of the graphene underneath it.

Figure 2 shows larger-scale AFM images together with LEEM results for the same samples pictured in Fig. 1. The LEEM images in Fig. 2(b) and (f) show data acquired with 3.7 eV electrons incident on the surface, with the measured signal being the reflectivity (*i.e.* bright-field imaging) of the electrons. As discussed by Hibino *et al.*, the reflectivity of electrons in the range 0 – 10 eV shows distinct oscillations arising from the existence of discrete energy levels in the conduction band of graphene with wave vectors normal to the surface. Each such state produces a minimum in the reflectance (maximum in the transmission), and for an *n*-ML thick film there are *n* such minima.[16] Reflectivity curves as a function of energy are shown in Figs. 2(c) and (g), displaying these oscillations.

From sequences of images acquired with an energy spacing of 0.1 eV, we analyze the reflectivity data to obtain local graphene coverage as follows: (i) at each pixel a reflectivity curve extending between about 2.0 and 6.5 eV is extracted from the data (ii) a quadratic background is subtracted, (iii) a sinusoidal function with adjustable frequency and phase is fit to the curve, and (iv) a scatterplot of the phase vs. frequency is viewed, with curves associated with different numbers of ML occupying distinctly different regions in the plot. This method of analysis is found to work well even when the low energy (≲1.7 eV) secondary electron peak obscures some local minima of a reflectivity curve. Figure 3 shows representative scatterplots of the reflectivity data, Fig. 3(a) corresponding to the sample of Fig. 2(b), and Fig. 3(b) to the sample discussed in Section III(C). On each scatterplot, the regions of frequency and phase corresponding to integer ML thicknesses are apparent, and from those we generate color-coded maps of graphene thickness.

For the relatively short annealing time of Fig. 2(b), the surface is found to be covered mainly with 2 ML of graphene, as seen in the map of Fig. 2(d). Not all of the 2 ML domains are equivalent, however, since their contrast in Fig. 2(b) appears to be quite mottled (a LEEM image from a similar film has recently been presented by Riedl et al.[20]). This mottling arises from variation in the magnitude of the intensity maximum at 3.7 eV, as illustrated by curves B, C and D of Fig. 2(c). We have not investigated this intensity variation further, but presumably it is related to the nm-scale nature of the small graphene domains in this nucleation phase of the film formation. In any case, Fig. 2(d) also shows that, in addition to the 2 ML areas, there are few small regions with 1 or 3 ML of graphene coverage.

With longer annealing time, the graphitized Si-face morphology undergoes considerable changes, as shown in Fig. 2(e). Step bunches form, one of which is marked by the arrows in Fig. 2(e). The color-coded map of the thickness is shown in Fig. 2(h), with this surface area having an average graphene thickness of 3.0 ML. Importantly, the thickest regions of graphene are found near step bunches, one of which extends in a zig-zag manner from top to bottom of Fig. 2(f) and Fig. 2(h). Away from the step bunches, the graphene thickness is predominantly 2 or 3 ML. Additional color-coded images of graphene thickness for this same surface are shown in Fig. 4. We associate all of the 4 or 5 ML regions with steps or step bunches on the surface, and we find a separation between bunches of ≥10 μm. On the flat terraces between step bunches we find that almost the entire surface has only 2 or 3 ML of graphene thicknesses. This type of thickness variation, restricted to a single ML, is indicative of layer-by-layer formation of the graphene on the areas between bunches. The occurrence of layer-by-layer graphene formation was also reported recently in a LEEM study by Hibino et al. of graphene on vicinal SiC surfaces.[21]

**B. Disordered Secondary Surface Phase**

Many AFM images of our Si-face graphene sample display a faint "snowflake-like" pattern, as mentioned above in regard to Fig. 1(b). This pattern is suggestive of some sort of secondary phase other than regular, ordered graphene. Previously we speculated that this phase might be associated with excess Si and/or C atoms present at the graphene/SiC interface,[14] but here we present additional data indicating that this phase probably exists on the top *surface* of the graphene. In this regard, we point out our recent study of graphene on the SiC($000\bar{1}$) surfaces (the so-called C-face) where we also observe the presence of a disordered surface phase for certain samples.[22] That surface phase was found to be composed purely of carbon, and from an observed shift in the G line in Raman spectroscopy we surmised that it most likely is NCG.

Below we argue that the disordered secondary phase that we observe on the Si-face is also NCG, and we discuss possible origins of this carbonaceous surface layer.

Figure 5 shows AFM and STM images obtained from a Si-face sample that was annealed in vacuum at 1350°C for 40 min. This sample was found from AES measurements to be covered with an average of 6.2 ML of graphene (a remarkably thick layer, as further discussed below). As seen in Fig. 5(a) there are two types of surface regions, the one labeled A is relatively flat and smooth whereas the other labeled B is rougher and has higher topographic height than A. An STM image of a flat, smooth area is shown in Fig. 5(b). The particular pattern found there can be definitely identified as the $6\sqrt{3}\times6\sqrt{3}$-R30° structure associated with the C-rich SiC(0001) surface;[23] this same pattern is found by STM imaging of other SiC(0001) surfaces, *both with or without* the presence of overlying graphene layers. Furthermore, from low-bias imaging, Fig. 5(c), a hexagonal corrugation with period consistent with that of graphene, $\sqrt{3}(0.246$ nm$)/2=0.213$ nm, is seen (the two dominant corrugations are marked by black lines in Fig. 5(c), with the third corrugation being very faint in the image). From this hexagonal arrangement we can identify the presence of graphene covering the $6\sqrt{3}\times6\sqrt{3}$-R30° structure, *i.e.* the usual surface phase for vacuum-annealed SiC(0001) surfaces.

An STM image over a larger range is shown in the upper part of Fig. 6, now including both an area of the graphene-covered $6\sqrt{3}\times6\sqrt{3}$-R30° structure (on the left) and a portion of the rough, disordered region (on the right). Tunneling spectra acquired at the points indicated are shown below the image. The spectra from the graphene-covered $6\sqrt{3}\times6\sqrt{3}$-R30° region are very similar to each other, and they closely resemble that known for graphene on SiC.[23] The spectra reveal nonzero conductance at 0 V, but nevertheless with a distinct conductance minimum at 0 V, indicative of semi-metallic behavior (these measured spectra actually have a somewhat lower conductance at 0 V than that found for single- or bilayer-graphene,[23] but in measurements of very thick graphene layers, *i.e.* graphite, we have observed similar low conductance at 0 V and we associate it with the presence of the semi-metallic graphitic band structure[24]). On the disordered region the *average* spectrum is quite similar to that from the graphene, but we find significant variation from point to point. Some points display a large conductance maximum near −1.1 or −0.5 V, others shown distinct features at positive voltages, and some curves are smoothly varying without any clear peaks. Although it is certainly not possible to identify the chemical composition of the disordered layer based solely on these tunneling spectra, we can nevertheless conclude that this material has an average spectrum quite similar to graphene but with additional features at specific locations, perhaps arising from extra dangling bonds and/or defects.

The fact that the spectra acquired on the disordered region can be, locally, quite different than that from the ordered graphene region indicates that the disordered region does not exist at the *interface* of the graphene and the SiC (since a tunneling spectrum of that would be smoothly varying, like the spectrum from the ordered region, at least for ≥2 ML of overlying graphene[23]). Rather it appears that the disordered material exists on the graphene *surface*. Cross-sectional cuts from the image of Fig. 5(a) are shown in Fig. 7. The disordered layer can be seen to have two different levels [see e.g. the arrow in Fig. 5(a) showing a transition between the levels], and from Fig. 7 it appears that the height of each level is about 0.37±0.02 nm, close to the half-unit-cell height of 6H-SiC, 0.375 nm. This latter agreement is coincidental, however, since from measurements made on 4H-SiC (not shown) we find a value for the height of the disordered layer of 0.35±0.03 nm, in agreement with the results from 6H-SiC.

We interpret the disordered surface layer on this Si-face sample, as well as the prior snowflake-like patterns seen on other Si-face samples,[14] in the same way as the disordered layers seen previously on C-face SiC, namely, as being NCG. The tunneling spectra we observe in Fig. 6 are consistent with this interpretation. The layer heights just mentioned are slightly larger than the expected 0.34 nm for NCG, but still close to that. AES results for the sample of Fig. 5 produced an intensity ratio of the C KLL line to the Si KLL line of 130. The intensity of the Si peak in this case was very low (almost unobservable), indicative of little or no Si content in the disordered surface layer. Taking the Si content to be zero, then the observed intensity ratio corresponds to a graphene+NCG coverage of 6.2 ML. This coverage is very high compared to other samples prepared under similar conditions; usually for a temperature of 1350°C we would achieve ≤2 ML of graphene on the surface (*e.g.* as in Fig. 1(a)).

We interpret the unusually high graphene coverage for the sample of Fig. 5, as well the presence of the disordered surface layer itself, in the following manner. First, we recall from our prior studies of NCG on the C-face that it is only certain wafers that produce a lot of NCG on their surfaces, with other wafers producing little.[22] The former types of wafers were found to have on their surfaces, after H-etching, a relatively large number of spiral step structures (density $>1\times10^4$ cm$^{-2}$), indicative of a high dislocation density in these wafers. Thus, we believe that it is the dislocation (and/or micro-pipe) density that is the main variable between the wafers. Perhaps it is the spiral step structures themselves that lead to the NCG, through decomposition of the curved step edges. Alternatively, the dislocation cores could perhaps act as a source of carbon (*e.g.* perhaps related to the discussion in Ref. [25]). In situations where large amounts of this carbon are produced, it apparently forms the NCG on the surface. Furthermore, we also consider it likely that this NCG could itself, at sufficiently high temperature, convert to ordered graphene. That conversion process would account for the exceptionally large graphene coverage on the sample of Fig. 5.

We note finally that the presence of the disordered, NCG layers, on the Si-face or the C-face, is found *not* to interfere with LEEM imaging of graphene layers beneath the NCG. The sample of Fig. 1(b), for example, revealed in its LEEM images [*e.g.* Fig. 2(f)] no significant trace of the snowflake-like pattern; rather, the graphene domains were nicely continuous on the length scale of the snowflake-like patterns. With sufficiently thick NCG (*e.g.* as occur on the C-face) then the overall intensity of the reflectivity is of course affected, but other than that we find the same reflectance oscillations (and diffraction pattern) from areas covered with NCG as from neighboring areas. We conclude, consistent with the discussion in the preceding paragraphs, that the NCG is simply sitting on top of an otherwise well ordered graphene layer.

**C. Graphene Formation in Argon**

As discussed in Section III(A), vacuum annealing of the Si-face SiC is found to found to form graphene in predominantly in a layer-by-layer manner, at least for thicknesses >2 ML. However, for graphene films thinner than 2 ML, *i.e.* for monolayer graphene in particular, we do not achieve layer-by-layer formation, as illustrated in Figs. 2(a)-(d). Of course it is desirable to achieve layer-by-layer growth (and hence the possibility of precise integer ML coverage) of the graphene for all possible coverage. How can this be achieved for monolayer graphene? The solution has recently been described by two research groups, who have both performed the annealing of the SiC in an environment of 1 atm argon.[9,10] In the presence of argon, the sublimation rate of Si from the SiC is reduced, and hence the annealing temperature needed for a given graphene thickness is increased significantly compared with the temperature used for a vacuum annealed film. This increased temperature leads to improved kinetics in the graphene formation process. Use of disilane, at $10^{-8} - 10^{-6}$ Torr, has been found to have a similar effect.[11]

We have succeeded in forming a single uniform monolayer of graphene on the Si-face of SiC by annealing in 1 atm argon, as shown in Fig. 8. This sample is annealed at a temperature of ≈1600°C for 30 min in a 1-atm environment of argon (99.999% purity). Prior to the argon annealing, the sample is H-etched which gives rise to an ordered array of steps and terraces on the surface. As a result of the argon annealing the steps undergo considerable motion, and we see in the AFM image of Fig. 8(a) large flat terraces separated by step bunches. A LEEM image of this sample, acquired at 2.8 eV, is shown in Fig. 8(b). Reflectivity curves from areas marked A – E in the LEEM image show only a single minimum and the corresponding scatterplot, shown in Fig. 3(b), exhibits only a single peak thus demonstrating that the surface is covered with graphene of monolayer thickness. Even though the surface is found to be covered with a single layer of graphene, we nevertheless see some contrast in the LEEM image of Fig. 8(b). This contrast might arise due to slight phase difference in the reflected electrons leading to a variation in their intensity.[26] LEEM images from other regions on the surface revealed some areas that were covered with bilayer graphene, but those accounted for only a few % of the total surface area.

In contrast to the results of Fig. 8, we observe for vacuum annealed samples of thickness ≈1 ML many small pits on the surface [*i.e.* similar to Fig. 2(a)]. Faster nucleation of the buffer layer at the higher annealing temperatures employed in argon is believed to inhibit this pit formation.[9,15] We do, however, observe a few relatively large pits on the surface for samples prepared in argon, as seen in Fig. 8(a). These pits could, in principle, arise from coarsening of smaller pits that occurred earlier in the graphene nucleation phase. Alternatively, the large pits of Fig. 8 could have a different origin, such as preferential sublimation at dislocations. Further studies are needed to clarify this point. In any case, compared with a vacuum annealed sample of thin coverage, such as Figs. 2(b) and (d), the samples prepared in argon are found to have much larger domains of uniform graphene thickness.

**SUMMARY**

We have studied the effect of annealing environment on the morphology and thickness of graphene prepared on SiC(0001) substrates. For films prepared by vacuum annealing, we observe areas of different thicknesses of graphene on the surface, with the lateral extent of these areas being about 3 – 4 μm. The surface shows relatively thick graphene near step bunches, but away from steps the graphene is thinner and is found to grow predominantly in a layer-by-layer mode. We investigated the presence of a secondary phase, which reveals itself as snowflake-like pattern in the morphology, using STM/STS. We believe that this phase consists of excess carbon which fails to form ordered graphene, and remains in a disordered form on the surface. This disordered surface phase, however, does *not* significantly perturb the LEEM reflectivity oscillations of the graphene lying beneath it. In order to achieve more homogenous graphene films (particularly for 1 ML thickness), samples have been annealed at higher temperatures in the presence of argon. The resulting film is found to have domains of 1-ML-thick graphene with domain size $\gg 10$ μm.

**ACKNOWLEDGEMENTS**
This work was supported by the National Science Foundation, grant DMR-0856240. Discussions with Gong Gu and Guowei He are gratefully acknowledged.

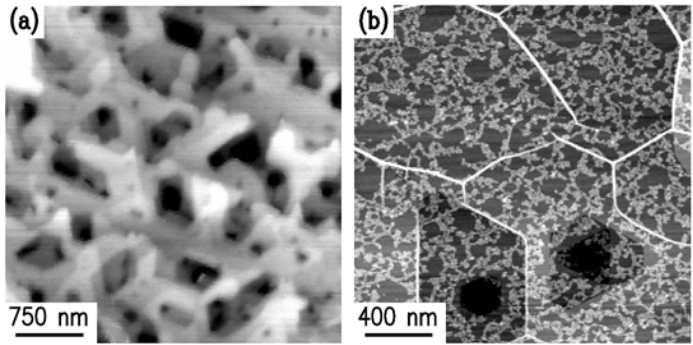

FIG 1. AFM images of graphene on (a) 6H-SiC(0001) and (b) 4H-SiC(0001), prepared by annealing in vacuum at 1320°C for (a) 10 min, and (b) 40 min. The resulting graphene coverages are 1.9 and 3.0 ML, respectively. Images are displayed with gray scale ranges of 3.0 and 1.8 nm, respectively.

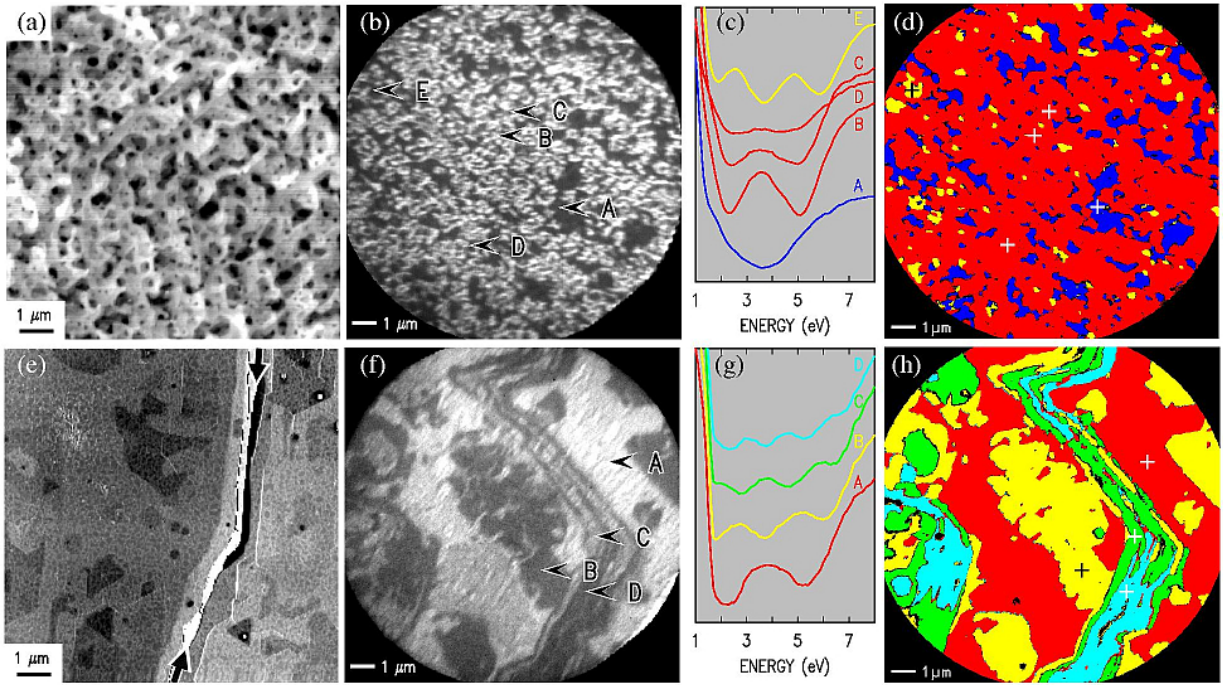

FIG 2. (Color online) AFM and LEEM results for graphene on SiC(0001), with (a)-(d) showing the same sample as Fig. 1(a) and (e)-(h) showing the same sample of Fig. 1(b). (a) and (e) AFM images, with (a) displayed using a gray scale range of 3 nm. For image (e) there is an 8-nm-high step bunch extending vertically across the image, indicated by the arrows, and hence a split gray-scale is used with 4 nm range for each of the terraces on either side of the step bunch. (b) and (f) LEEM images at an electron beam energy of 3.7 eV. (c) and (g) Intensity of the reflected electrons from different regions marked in (b) or (f) as a function of electron beam energy. (d) and (h) Color-coded maps of local graphene thickness, deduced from analysis of the intensity vs. energy at each pixel; blue, red, yellow, green, and cyan correspond to 1, 2, 3, 4, and 5 ML of graphene, respectively. Small white or black crosses mark the locations of the intensity vs. energy curves. Regions with no discernable oscillations are colored black.

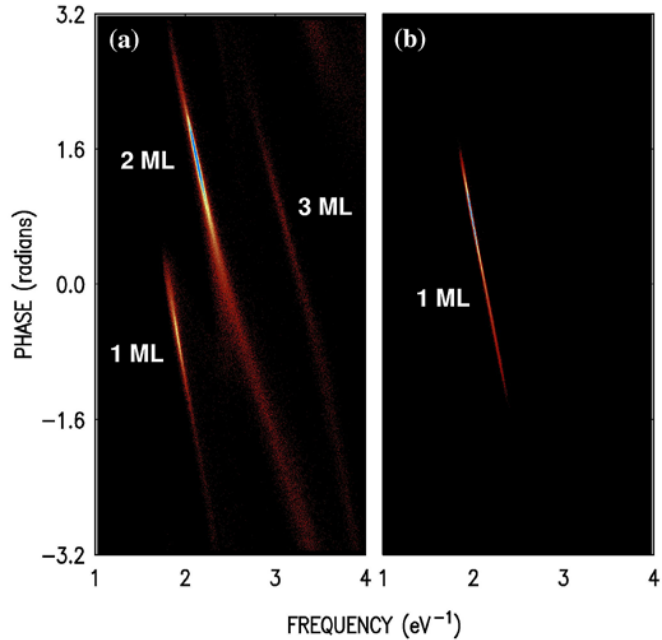

FIG 3. (Color online) Scatterplots of phase vs. frequency, resulting from analysis of reflectivity curves by fitting them to a function $\sin(\alpha E + \phi)$ where $E$ is the electron energy, $\alpha$ is the frequency, and $\phi$ is the phase. Each scatterplot comes from analysis of a sequence of LEEM images ranging between 2.0 and 6.5 eV. Panels (a) and (b) show the results of analysis of LEEM data for the samples of Fig. 2(b) and Fig. 8(b), respectively.

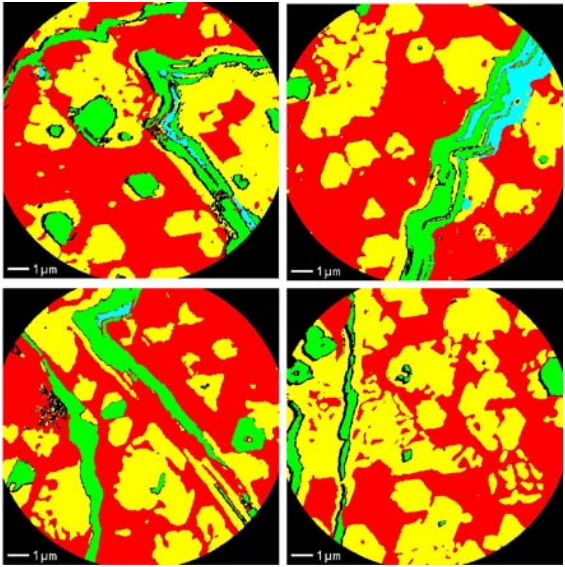

FIG 4. (Color online) Color-coded maps of local graphene thickness for the same sample (and with the same manner of display) as Fig. 2(h).

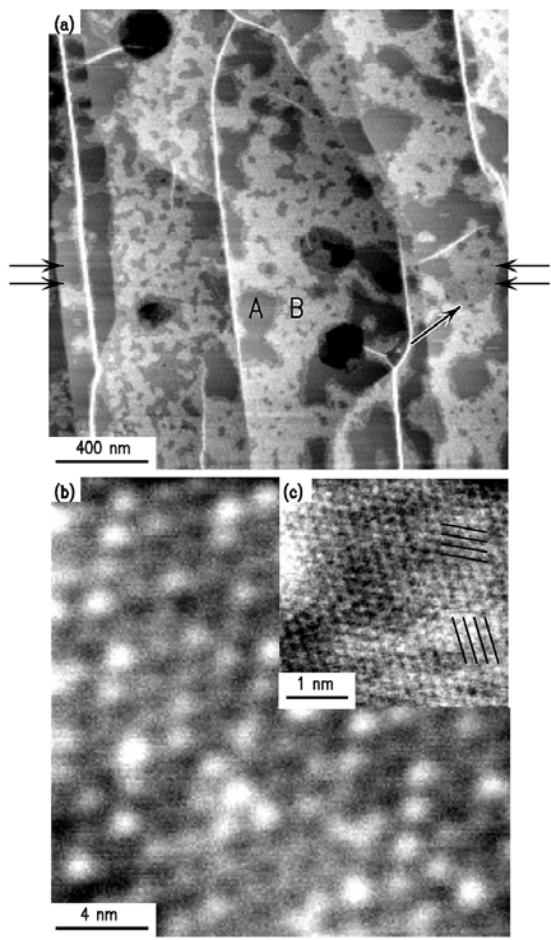

FIG 5. Images of graphene on SiC(0001), prepared by annealing in vacuum at 1350°C for 40 min. (a) AFM image, (b) STM image acquired at sample voltage of +2.0 V, and (c) STM image acquired at +0.4 V. Images are displayed with gray scale ranges of 2.2, 0.04 nm, and 0.025 nm, respectively. Arrows on the side of (a) indicate the locations of cross-sectional cuts, shown in Fig. 7. Black lines in (c) indicate the two dominant corrugations in the image.

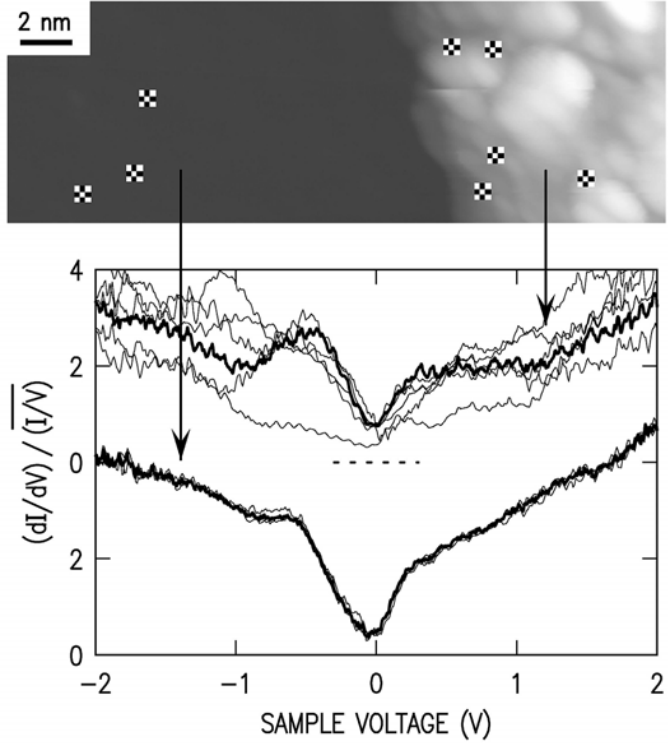

FIG 6. Upper: STM image acquired from same sample as in Fig. 5, with a sample voltage of +2.0 V and displayed with gray scale range of 2.2 nm. Lower: Tunneling spectra from the locations indicated, with individual spectra are shown by thin lines and the average by a thick line.

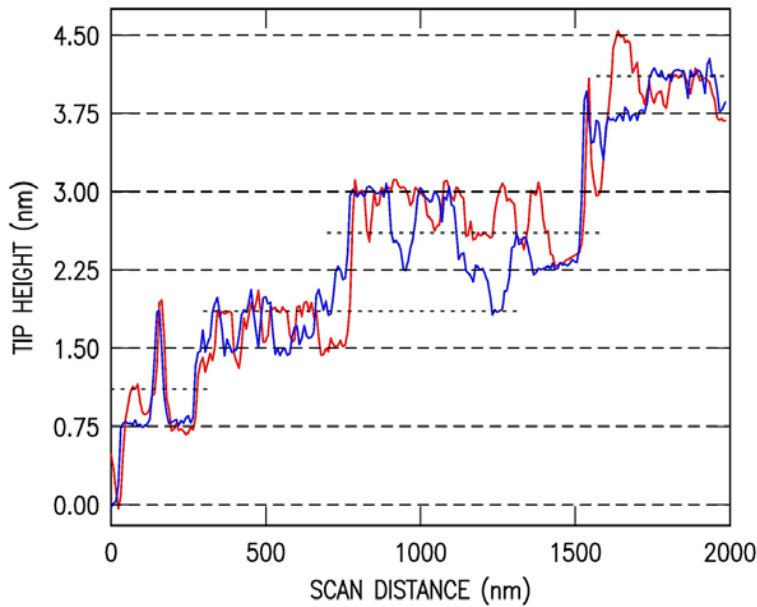

FIG 7. (Color online) Topographic height profiles at the locations of the arrows in the AFM image of Fig. 5(a), with the blue (red) line corresponding to the upper (lower) cut. Dashed lines in the surface height indicate intervals of 0.75 nm, the half-unit-cell height of 6H-SiC(0001), with dotted lines showing half that amount.

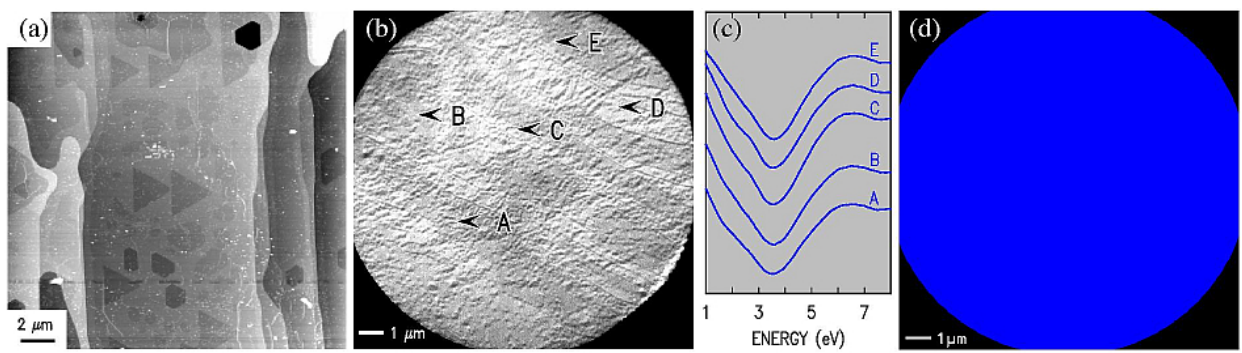

FIG 8. (Color online) Results of graphene prepared on SiC(0001), by annealing at 1600°C for 30 min in 1 atm of argon. (a) AFM image with gray scale range of 8 nm, (b) LEEM image at beam energy of 2.8 eV, (c) Intensity of the reflected electrons from different regions marked in (b) as a function of electron beam energy, and (d) Color-coded maps of local graphene thickness, deduced from analysis of the intensity vs. energy at each pixel; blue corresponds to 1 ML. The surface is covered uniformly with monolayer thick graphene.